\documentclass{sigchi}

\usepackage{balance}
\usepackage{graphics}
\usepackage{times}
\usepackage{url}
\usepackage{booktabs}

\makeatletter
\def\url@leostyle{%
  \@ifundefined{selectfont}{\def\UrlFont{\sf}}{\def\UrlFont{\small\bf\ttfamily}}}
\makeatother
\def\pprw{8.5in}
\def\pprh{11in}

\usepackage[pdftex]{hyperref}
\hypersetup{
pdftitle={Understanding Chatbot-mediated Task Management},
pdfauthor={LaTeX},
pdfkeywords={SIGCHI, proceedings, archival format},
bookmarksnumbered,
pdfstartview={FitH},
colorlinks,
citecolor=black,
filecolor=black,
linkcolor=black,
urlcolor=black,
breaklinks=true,
}

\toappear{\scriptsize This work is made available under the terms of the 
Creative Commons Attribution 4.0 International License (CC-BY- 4.0)}

\begin{document}

\title{Understanding Chatbot-mediated Task Management}

\numberofauthors{3}
\author{
  \alignauthor Carlos Toxtli\\
    \affaddr{West Virginia University}\\
    \affaddr{Morgantown, WV, USA}\\
  \alignauthor Andr\'{e}s Monroy-Hern\'{a}ndez\thanks{Currently affiliated with Snap Inc.}\\
    \affaddr{Microsoft Research}\\
    \affaddr{Redmond, WA, USA}\\
  \alignauthor Justin Cranshaw\\
    \affaddr{Microsoft Research}\\
    \affaddr{Redmond, WA, USA}\\
}

\maketitle

\begin{abstract}
Effective task management is essential to successful team collaboration. While the past decade has seen considerable innovation in systems that track and manage group tasks, these innovations have typically been outside of the principal communication channels: email, instant messenger, and group chat. Teams formulate, discuss, refine, assign, and track the progress of their collaborative tasks over electronic communication channels, yet they must leave these channels to update their task-tracking tools, creating a source of friction and inefficiency. To address this problem, we explore how bots might be used to mediate task management for individuals and teams. We deploy a prototype bot to eight different teams of information workers to help them create, assign, and keep track of tasks, all within their main communication channel. We derived seven insights for the design of future bots for coordinating work.
\end{abstract}

\keywords{
	Conversational User Interfaces; chatbots; bots; productivity; tasks; 
}

\category{H.5.m.}{Information Interfaces and Presentation (e.g. HCI)}{Miscellaneous}

\section{Introduction}

Information workers are constantly switching between communication and productivity tools \cite{czerwinski2004diary, iqbal2010notifications}. For example, people might have an email conversation to decide on a time to meet, then switch to their calendaring tool to add the meeting to their schedule. Similarly, teams might discuss a project over Slack, decide who gets to work on what, then add those tasks to a task management system. These interruptions can add up to reduced productivity and increased stress in the workplace \cite{czerwinski2000instant, iqbal2007disruption, mark2008cost}.

In order to reduce the cost of switching contexts, a new category of productivity technologies have emerged in the form of chatbots that can be summoned from within communication channels. People can delegate tasks in-situ to these chatbots, or bots for short, without having to leave the chatroom, messenger app, or email client. For example, a scheduling meeting bot service allows people to delegate the work of scheduling a meeting by cc'ing the bot in an email conversation \cite{cranshaw2017calendar}. Similarly, there is a myriad of bots available in work-centric chat platforms like Slack and Microsoft Teams, from bots that help teams order food \cite{kip}, to those that integrate with software development tools \cite{lin2016developers}. 

Although technologies for team coordination of tasks \cite{Sanchez:2008:ITC:1378773.1378850, Anh:2012:DCP:2372251.2372274} are well-established, the emergence of bots as mediators of work-related activities presents new challenges and opportunities, some of which we investigate in this work.

\begin{figure}
    \centering
    \includegraphics[width=\columnwidth]{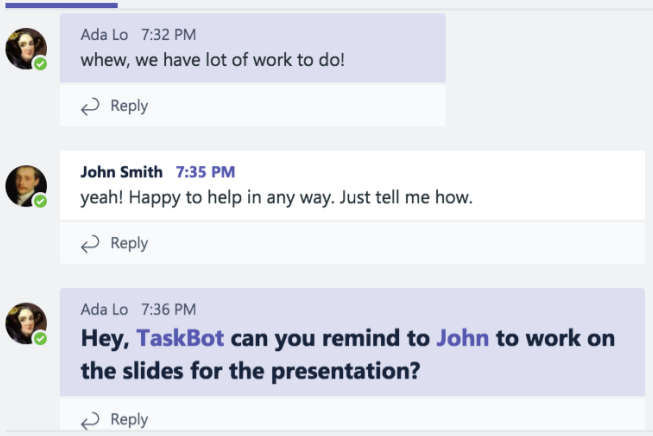}
    \caption{A group chat conversation on \textit{Microsoft Teams} showing a user assigning a task to her teammate and asking TaskBot to help track of it.}
    \label{fig:screenshot}
\end{figure}

In this work, we focus on trying to understand how bots can mediate task management through the iterative design, engineering, and deployment one such bot: TaskBot. Members of a team can, for example, interact with TaskBot by mentioning it in a conversation in the team's chatroom an delegating to the bot the job of tracking and reminding people to complete their tasks (see Figure \ref{fig:screenshot}).

We deployed TaskBot in eight different teams of information workers, who used TaskBot to help them track 88 distinct tasks. We present several novel findings on how people used TaskBot to mediate work may be useful to future designers, including how people use TaskBot to assign tasks to others, how created self-reminders with TaskBot, how people treated TaskBot in human-like ways, and how it is a challenge to managing multiple tasks simultaneously.

\begin{table}[t!]
  \centering \small
  \begin{tabular}{p{2.2cm} p{6.0cm} }
    \toprule
   {\bf Intent}&{\bf Examples}\\\midrule
{\bf Deadline Setting} 
& \emph{I'll be done tomorrow}, or \emph{Next week}, or \emph{Thursday}.\\
{\bf Task Completion}  
& \emph{I'm done already}, or \emph{It's finished}, or \emph{Yes}.\\
{\bf Task Rejection}  
& \emph{Cancel the task}, or \emph{Unassign}, or \emph{No}.\\  \bottomrule
  \end{tabular}
  \caption{Reactive messages intents in response to ``Are you done with the task? If not, when do you think you'll finish?''}
  \label{table:reactive}
\end{table}

\section{Related work}

Chatbots began attracting attention in the formative days of artificial intelligence research \cite{shawar2007chatbots} as a vehicle for demonstrating machine understanding \cite{turing1950computing}. While early experiments focused on free-form conversations \cite{weizenbaum1966eliza}, bots have recently become more popular as single purpose tools, performing a complex tasks for people via a conversational interface \cite{cranshaw2017calendar, fast2017iris}. We build on this more utilitarian take on bots, offering a convenient way for teams to track and manage their tasks from within a conversation.

Previous research has looked at the intersection between task tracking and communication. Some systems seek to extend or enhance the email client with extra capabilities to track tasks \cite{bellotti2003taking, bellotti2005quality, whittaker2005supporting}. Our work differs from this approach, in that it does not rely on any particular messaging client, making it more inter-operable with the tools and workflows of peoples' choosing. Other work has looked at how to proactively identify tasks in messages \cite{faulring2010agent, yang2017characterizing, sappelli2016assessing, corston2004task}, and for example, classify them and present them in a task-centric interface from where users could easily find people who had performed that task before \cite{faulring2010agent}. While these seamless, proactive approaches show promise, in practice, it is difficult to accurately detecting the users' precise intentions about a task, so we opt to give them control over when and how they invoke the bot.

A number of products exist for collaborative task tracking for teams, including Trello~\cite{trello}, Asana~\cite{asana}, Wunderlist~\cite{wunderlist}, and Visual Studio Online~\cite{vso}. While these products vary considerably in terms of offerings, generally they allow teams to create tasks records, add implementation and execution details, assign tasks to team members, track the task progress, and develop time-lines or plans of execution for pending work. TaskBot extends this basic approach, offering a conversational interface to the core functionality of these tools, using Wunderlist as its backend. 
Some companies have begun exploring bots for tracking task, including To-do bot for slack \cite{Todobotb64:online} and the Trello bot for Microsoft Teams \cite{TheTrell83:online}, however, these bots have taken an command-driven approach to building a conversational user experience, rather than focusing on free form conversational interactions, as does TaskBot.

\section{TaskBot User experience}

TaskBot helps teams manage their tasks from within platforms they use to communicate. Although TaskBot is available on several different platforms, such as email, Skype for Business, and Slack, in this work we focus on Microsoft Teams, a group chat platform for the workplace. 

Once an administrator adds TaskBot to a chatroom, anyone on the team can then interact with it by @-mentioning ``@TaskBot" and some of their teammates in a message. For example, Alice might summon TaskBot to help her assign a task to her colleague Bob and follow up with him until he completes the it:

(1) Alice asks the bot to help her track a task for Bob. 

\begin{quote}
\small
\texttt{Alice: @TaskBot Please remind @Bob to submit the proposal.}

\texttt{TaskBot: Okay, I got your request.}

\end{quote}

 If Alice had mentioned two or more people in her message, then TaskBot assigns the task to all mentioned parties. We settled on this approach after trying and discarding, other methods, including a round-robin approach where TaskBot first asked each person if they could work on the task.
 
(2) Next, TaskBot asks Bob for an ETA for Alice's request.
\begin{quote}
\small

\texttt{TaskBot: Hi Bob, Alice recently asked you to take this task:  "Please remind Bob to submit the proposal." When do you think you can have it done?.}

\texttt{Bob: Tomorrow.}

\texttt{TaskBot: Bot: Okay, got it. I'll check in with you then.}
\end{quote}

Because these conversations are one-on-one with the bot, they do not add noise to the to the rest of the team's conversations:

(3) The next day, TaskBot checks in with Bob:
\begin{quote}
\small

\texttt{TaskBot: Are you done with "Please remind Bob to submit the proposal."? If not, when you think you'll finish?.}

\texttt{Bob: I'm done.}

\texttt{TaskBot: Great job, Bob! I marked this task as complete.}
\end{quote}

This adaptive approach to following up at the specific time was an improvement over  earlier versions of the system, where TaskBot checked on the tasks twice a day (in the morning and in the afternoon), which early pilot participants considered too frequent and annoying.

(4) Finally, TaskBot let's Alice know Bob is done.

\begin{quote}
\small
\texttt{TaskBot: Good news! Bob completed the task: "Please remind Bob to submit the proposal."}
\end{quote}

\begin{table}[t!]
  \centering \small
    \begin{tabular}{p{2.4cm} p{6.0cm} }
    \toprule
   {\bf Intent}&{\bf Example}\\ \midrule
{\bf Task assignment} 
& \emph{@Bob, don't forget to finish the report. @TaskBot}\\
{\bf Self-Reminders}  
& \emph{@TaskBot, please remind me to leave at 5pm.}\\
{\bf Greetings}  
& \emph{Thanks, TaskBot!}\\
{\bf Task Termination}  
& \emph{I've completed this task,} or \emph{Please cancel.}\\
{\bf Help}  
& \emph{Help me!}\\
{\bf Other}  
& Catch-all: TaskBot replies with ``I didn't understand." \\\bottomrule
  \end{tabular}
  \caption{Proactive message intents to new requests.}
  \label{table:proactive}
\end{table}

\section{Implementation}

We built TaskBot using the Microsoft Bot Framework's Bot Builder SDK for Node.js. The state of the tasks were stored in Wunderlist and accessed by TaskBot via the Wunderlist API. For this study, all tasks were stored under a central Wunderlist account for which the bot was the owner, but in practice, we envision people integrating TaskBot with their own accounts on task tracking platforms of their choosing.

\begin{figure}[t]
    \centering
    \includegraphics[width=\columnwidth]{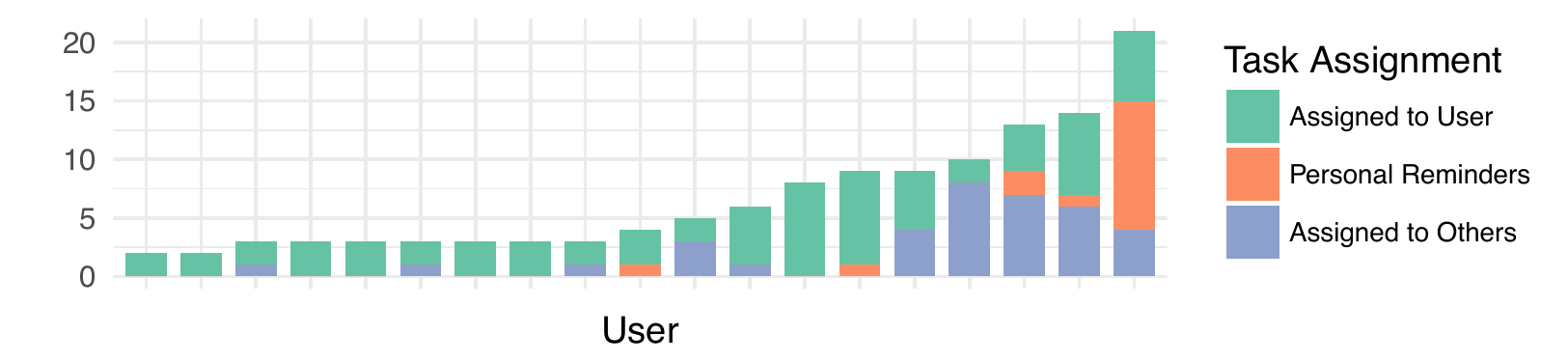}
    \caption{Usage statistics of 19 users in the study, by assignment type, showing number of tasks assigned to the user by others, assigned to others by the user, and personal reminders (tasks assigned to one's self).}
    \label{fig:usage}
\end{figure}

We implemented a model for detecting intents from users' utterances using the Microsoft Language Understanding Intelligent Service (LUIS), focusing on two possible categories of messages TaskBot receives from a user:
\begin{enumerate}
    \item Reactive messages in response to a key TaskBot's question: ``Are you done with the task? If not, when do you think you'll finish?'' (see Table \ref{table:reactive}).
    \item Proactive messages, that is, unprompted messages people post in a group chat mentioning TaskBot, or send to TaskBot via direct message (see Table \ref{table:proactive}).   
\end{enumerate} 
 We trained this model by hand at first, based on lists of keywords we developed to capture these intents, and then we iteratively improved it with users' actual utterances in several rounds of pilot testing, first within our own team, then in our extended research group.

\section{Methods}

We conducted a week-long field deployment of TaskBot with eight teams of information workers ranging in size from two to five people, with 19 people in total. Participants were employees from the Microsoft Redmond campus, who were recruited via email using company mailing lists. Five of the eight teams were hierarchical, consisting of a manager and one or more of their subordinates, while the others were peers at the same level. All teams were collocated, and communicated via synchronous, asynchronous, and face-to-face interactions.

A requirement for participation was that the teams of two or more people had to be working closely together on the same project, and all members of the team had to agree to be  part of the study. Each team was given a brief tutorial on how to use TaskBot, including the creation of an example task. Over the course of one work week, each participant was asked to create at least three tasks using TaskBot. We also administered pre- and post- study surveys. In gratuity, each participant received a \$20 gift card.

After the field study, all the messages written by users to TaskBot (177) were independently coded by two of the authors into 27 categories, that were inductively and iteratively developed to characterize usage behavior. The two independent classifications agreed on 95.6\% of all the messages (Cohen's kappa = .76). The remaining discrepancies were resolved through aggregation or discussion as appropriate.

\section{Results}

Over the course of the field deployment, participants created 88 tasks---averaging 4.4 tasks per user. People assigned to those tasks marked 65 of them as ``complete.'' Figure \ref{fig:usage} shows how these tasks are distributed across users and types of task assignment. Table \ref{table:usageperteam} shows how they are distributed across teams. The median number of tasks created per team was seven.

\begin{table}[t!]
  \centering \small
    \begin{tabular}{p{1.2cm} p{0.5cm} p{0.5cm} p{0.5cm} p{0.5cm} p{0.5cm} p{0.5cm} p{0.5cm} p{0.5cm}}
    \toprule
   {\bf Teams}&{\bf 1}&{\bf 2}&{\bf 3}&{\bf 4}&{\bf 5}&{\bf 6}&{\bf 7}&{\bf 8}\\ \midrule
{\bf Members} 
& \emph{3} & \emph{2} & \emph{3} & \emph{3} & \emph{2} & \emph{2} & \emph{2} & \emph{5}\\
{\bf Tasks}  
& \emph{20} & \emph{5} & \emph{12} & \emph{9} & \emph{20} & \emph{6} & \emph{5} & \emph{12}\\
{\bf Messages}  
& \emph{142} & \emph{70} & \emph{139} & \emph{131} & \emph{267} & \emph{54} & \emph{86} & \emph{170}\\\bottomrule
  \end{tabular}
  \caption{Usage per team.}
  \label{table:usageperteam}
\end{table}

In tracking the progress of these 88 tasks, TaskBot received 177 messages from participants. 22\% of these messages were tasks assignments. On average, users sent 12 messages to TaskBot, while users received 54 messages from TaskBot. Lastly, on average, users received 16 reminders to complete their assigned tasks. 

From the post-study survey, we learned that six of the 11 people who answered felt more productive, and 8 people will use it in the future, however, 4 people reported finding  it annoying. When noting features that were important to them, most said that reminding other people about tasks (10/11), and tracking the progress of tasks (9/11) as important. Six people did not find it important that TaskBot to interface with external task tracking software. 

\section{Discussion}

In this section, we highlight seven different patterns we observed from people's interactions with TaskBot
that are relevant lessons for designing work-centric chat bots.\footnote{These findings also align with our prior experience building productivity bots \cite{cranshaw2017calendar}.}

Users particularly enjoyed the ease with which they were able to create and track tasks from the conversation in-situ: ``\emph{I liked being able to jot tasks down and have them get tracked without needing to go to a separate tool to track them.}'' However, people expressed concerns about the complexity of using a conversational user interface for handling many tasks simultaneously: ``\emph{when I received multiple messages I didn't know how to respond to a specific message/reminder.}''  People also requested to use TaskBot on other channels instead of Microsoft Teams that they use more, for example, Skype for Business. Five people perceived reminders from TaskBot as invasive or annoying. Some cited the frequency of reminders as the main problem: ``\emph{The reminders were annoying and too frequent.}'' Others pointed to their lack of context: ``\emph{reminders are not context sensitive and would appear at random times and often caused me to task switch when I was focused in doing something else.}''

\subsection{Handling human-like interactions with bots}

People often treated TaskBot in a human-like manner, conversing with it naturally, and politely, using words like ``please" and ``thanks," which was somewhat unexpected. We coded all messages for signals of this type of behavior and we found that almost all the participants (93\%) interacted with TaskBot as if it was a human at least once.  In fact, some users (20\%) never interacted with TaskBot as if it was a bot. For example, one of the participants said:

\begin{quote}
\small
\tt{P1: Hi @TaskBot, could you remind @P2 to read the article I shared with him? Thanks!}
\end{quote}

Unfortunately, TaskBot sometimes failed to understand users when they interacted in polite or human-like ways, 
because the natural language understanding model hadn't been trained to handle those utterances, or they asked for something beyond its capabilities.

\emph{Designers of chatbots in the workplace should either, explicitly build signals that their bot is not personified, or be resilient to the range of possible responses when when people interact with the bot in more human ways.}

\subsection{Supporting self-communication}

Even though the user training focused on how to assign tasks to others, five users asked TaskBot to create reminders for themselves:

\begin{quote}
\small
\tt{P3: Remind me at 10:15 to leave}
\end{quote}

In some cases this was a way for people to get started without bothering others, but for some people it became a common practice, not unlike emailing oneself with notes and tasks.  

\emph{Designers of social chatbots should assume that bots would also be used for self-communication, either as a way to test the system or as a practical use of the tool.}

\subsection{Hierarchical task-assignment}

We observed people using TaskBot to assign tasks to people across different hierarchical levels. The percentage of  tasks that were assigned to managers by their subordinates (35\%), was almost the same as those assigned to subordinates by their managers (31\%). The rest (34\%) were among people at the same level.
When looking specifically at reminders of pending tasks (as opposed to assigning new tasks), we observed preliminary evidence that managers and subordinates used TaskBot differently.
For example, 83\% of requests that were upward in hierarchy (from subordinate to manager) were reminders, compared to only 47\% of tasks that were downward in hierarchy (from manager to subordinate).

\emph{Designers of social chatbots should expect different uses of the same bot based on people's hierarchy.}

\subsection{Failing gracefully}
TaskBot failed to understand about 10\% of messages people directed at it. This often happened because  participants did not know the range of features available, and they would try asking TaskBot to do something it cannot do. Also, because TaskBot was  not trained with a large enough corpus of data, otherwise simple utterances would fail to be understood.  In these circumstances, there were some messages (10.5\%) that TaskBot could not understand, and therefore were not answered successfully, for example two of the messages that TaskBot failed to understand were:

\begin{quote}
\small
\texttt{P7: How do I reply to a question?}
\end{quote}

\begin{quote}
\small
\texttt{P8: Can you send the consent form again? We've filled out the survey.}
\end{quote}

TaskBot's reply in these cases was a canned response, ``Sorry, I could not understand. Could you rephrase?''

\emph{Designers should create bots that fail gracefully when users ask for novel scenarios, and those failures should be categorized as such in order to use that data to improve future version of the bot. Designers should come up with mitigation strategies such as amusing error messages (e.g. Twitter's fail whale), and provide escalation to humans if possible \cite{cranshaw2017calendar}. }

\subsection{Dealing with human ambiguity}
Out of all the dates and times people mentioned in messages, 29\% of them were ambiguously defined, with 60\% of users mentioned at least one ambiguous date. For instance:

\begin{quote}
\small
\texttt{P3: Before tomorrow morning} 
\end{quote}

\emph{Sometimes people are intentionally ambiguous in conversations \cite{rong2017managing}, so designers of chatbots need to build strategies to resolve certain ambiguities to function according to user expectation.}

\subsection{Identifying people's name in conversations}
Group communication channels often use a specific syntax to mention people within the messages, e.g. the @ symbol. This helps bots like TaskBot  identify when someone is mentioned in a message. However, people didn't always use the special syntax and this created problems for TaskBot that was not able to understand that a string like ``Alice" was referring to a person without the prefix ``@Alice." About 40\% of the users experienced issues when they forgot to type the symbol "@" before the name. For instance, the TaskBot would fail to assign the following task to John:

\begin{quote}
\small
\texttt{P4: Hey John, can you finish your tutorial? cc @TaskBot}
\end{quote}

\emph{Designers should find ways of nudging users to mention people in the ways communication channels expect (e.g. using the at sign), or create smarter ways of detecting when a person might be mentioned in a message.}

\subsection{Handling multi-threaded conversations}

One of the biggest challenges for TaskBot and other bots is the difficulty of maintaining multiple active conversations at the same time. For example, some users had multiple tasks assigned to them, and when they told the bot ``I'm done with the task,'' the bot would not know which task they were referring to. We implemented a solution for this in TaskBot, using a menu for canceling and completing tasks that would list all active tasks. However, this was not the most natural or elegant interaction.

\emph{Designers should invest in technology for determining which active conversation thread a new incoming message belongs to.}

\section{Conclusions}

In this paper we introduce TaskBot, a bot designed to help teams manage their tasks. Users delegate the tracking of their tasks to TaskBot. We described our approach to designing TaskBot, and shared the lessons that we learned from deploying it with eight  teams. We focused on identifying design considerations for other bot designers building conversational user interfaces for workplace. As for TaskBot, future work will focus on the following features: exploring the use of multiple communication channels (e.g. email, Skype, etc.), better handling of multi-threaded conversations, and more sophisticated ways of assigning tasks to people based on the task description.

\balance

\bibliographystyle{acm-sigchi}
\bibliography{sample}

\begin{thebibliography}{10}

\bibitem{asana}
Asana.
\newblock \url{http://asana.com}.
\newblock (Accessed on 09/18/2017).

\bibitem{Todobotb64:online}
To-do bot by workast | slack app directory.
\newblock
  \url{https://cartheftbot.slack.com/apps/A0HBTUUPK-to-do-bot-by-workast}.
\newblock (Accessed on 09/18/2017).

\bibitem{trello}
Trello.
\newblock \url{http://trello.com}.
\newblock (Accessed on 09/18/2017).

\bibitem{TheTrell83:online}
The trello app for microsoft teams - trello help.
\newblock
  \url{http://help.trello.com/article/1086-the-trello-app-for-microsoft-teams}.
\newblock (Accessed on 09/18/2017).

\bibitem{vso}
Visual studio online.
\newblock \url{https://www.visualstudio.com/vso/}.
\newblock (Accessed on 09/18/2017).

\bibitem{wunderlist}
Wunderlist.
\newblock \url{http://wunderlist.com}.
\newblock (Accessed on 09/18/2017).

\bibitem{Anh:2012:DCP:2372251.2372274}
Anh, N.-D., Cruzes, D.~S., and Conradi, R.
\newblock Dispersion, coordination and performance in global software teams: A
  systematic review.
\newblock In {\em Proceedings of the ACM-IEEE International Symposium on
  Empirical Software Engineering and Measurement}, ESEM '12, ACM (New York, NY,
  USA, 2012), 129--138.

\bibitem{bellotti2003taking}
Bellotti, V., Ducheneaut, N., Howard, M., and Smith, I.
\newblock Taking email to task: the design and evaluation of a task management
  centered email tool.
\newblock In {\em Proceedings of the SIGCHI conference on Human factors in
  computing systems}, ACM (2003), 345--352.

\bibitem{bellotti2005quality}
Bellotti, V., Ducheneaut, N., Howard, M., Smith, I., and Grinter, R.~E.
\newblock Quality versus quantity: E-mail-centric task management and its
  relation with overload.
\newblock {\em Human-computer interaction 20}, 1 (2005), 89--138.

\bibitem{corston2004task}
Corston-Oliver, S., Ringger, E., Gamon, M., and Campbell, R.
\newblock Task-focused summarization of email.
\newblock In {\em ACL-04 Workshop: Text Summarization Branches Out} (2004),
  43--50.

\bibitem{cranshaw2017calendar}
Cranshaw, J., Elwany, E., Newman, T., Kocielnik, R., Yu, B., Soni, S., Teevan,
  J., and Monroy-Hern{\'a}ndez, A.
\newblock Calendar. help: Designing a workflow-based scheduling agent with
  humans in the loop.
\newblock In {\em Proceedings of the 2017 CHI Conference on Human Factors in
  Computing Systems}, ACM (2017), 2382--2393.

\bibitem{czerwinski2000instant}
Czerwinski, M., Cutrell, E., and Horvitz, E.
\newblock Instant messaging and interruption: Influence of task type on
  performance.
\newblock In {\em OZCHI 2000 conference proceedings}, vol.~356 (2000),
  361--367.

\bibitem{czerwinski2004diary}
Czerwinski, M., Horvitz, E., and Wilhite, S.
\newblock A diary study of task switching and interruptions.
\newblock In {\em Proceedings of the SIGCHI conference on Human factors in
  computing systems}, ACM (2004), 175--182.

\bibitem{fast2017iris}
Fast, E., Chen, B., Mendelsohn, J., Bassen, J., and Bernstein, M.
\newblock Iris: A conversational agent for complex tasks.
\newblock {\em arXiv preprint arXiv:1707.05015\/} (2017).

\bibitem{faulring2010agent}
Faulring, A., Myers, B., Mohnkern, K., Schmerl, B., Steinfeld, A., Zimmerman,
  J., Smailagic, A., Hansen, J., and Siewiorek, D.
\newblock Agent-assisted task management that reduces email overload.
\newblock In {\em Proceedings of the 15th international conference on
  Intelligent user interfaces}, ACM (2010), 61--70.

\bibitem{iqbal2007disruption}
Iqbal, S.~T., and Horvitz, E.
\newblock Disruption and recovery of computing tasks: field study, analysis,
  and directions.
\newblock In {\em Proceedings of the SIGCHI conference on Human factors in
  computing systems}, ACM (2007), 677--686.

\bibitem{iqbal2010notifications}
Iqbal, S.~T., and Horvitz, E.
\newblock Notifications and awareness: a field study of alert usage and
  preferences.
\newblock In {\em Proceedings of the 2010 ACM conference on Computer supported
  cooperative work}, ACM (2010), 27--30.

\bibitem{kip}
Kip.
\newblock What bots can do, that websites and apps can’t, mar 2017.

\bibitem{lin2016developers}
Lin, B., Zagalsky, A., Storey, M.-A., and Serebrenik, A.
\newblock Why developers are slacking off: Understanding how software teams use
  slack.
\newblock In {\em Proceedings of the 19th ACM Conference on Computer Supported
  Cooperative Work and Social Computing Companion}, ACM (2016), 333--336.

\bibitem{mark2008cost}
Mark, G., Gudith, D., and Klocke, U.
\newblock The cost of interrupted work: more speed and stress.
\newblock In {\em Proceedings of the SIGCHI conference on Human Factors in
  Computing Systems}, ACM (2008), 107--110.

\bibitem{rong2017managing}
Rong, X., Fourney, A., Brewer, R.~N., Morris, M.~R., and Bennett, P.~N.
\newblock Managing uncertainty in time expressions for virtual assistants.
\newblock In {\em Proceedings of the 2017 CHI Conference on Human Factors in
  Computing Systems}, ACM (2017), 568--579.

\bibitem{Sanchez:2008:ITC:1378773.1378850}
Sanchez, R., Jin, J., Maheswaran, R.~T., and Szekely, P.
\newblock Interfaces for team coordination.
\newblock In {\em Proceedings of the 13th International Conference on
  Intelligent User Interfaces}, IUI '08, ACM (New York, NY, USA, 2008),
  427--428.

\bibitem{sappelli2016assessing}
Sappelli, M., Pasi, G., Verberne, S., de~Boer, M., and Kraaij, W.
\newblock Assessing e-mail intent and tasks in e-mail messages.
\newblock {\em Information Sciences 358\/} (2016), 1--17.

\bibitem{shawar2007chatbots}
Shawar, B.~A., and Atwell, E.
\newblock Chatbots: are they really useful?
\newblock In {\em LDV Forum}, vol.~22 (2007), 29--49.

\bibitem{turing1950computing}
Turing, A.~M.
\newblock Computing machinery and intelligence.
\newblock {\em Mind 59}, 236 (1950), 433--460.

\bibitem{weizenbaum1966eliza}
Weizenbaum, J.
\newblock Eliza—a computer program for the study of natural language
  communication between man and machine.
\newblock {\em Communications of the ACM 9}, 1 (1966), 36--45.

\bibitem{whittaker2005supporting}
Whittaker, S.
\newblock Supporting collaborative task management in e-mail.
\newblock {\em Human--Computer Interaction 20}, 1-2 (2005), 49--88.

\bibitem{yang2017characterizing}
Yang, L., Dumais, S.~T., Benne, P.~N., and Awadallah, A.~H.
\newblock Characterizing and predicting enterprise email reply behavior.
\newblock In {\em Proceedings of the 40th International ACM SIGIR conference on
  Research and Development in Information Retrieval, SIGIR 2017, Tokyo, Japan}
  (2017).

\end{thebibliography}
\end{document}